\documentclass[aps,twocolumn,prc,amssymb,amsmath,superscriptaddress, bm, floatfix,10pt]{revtex4-1}
\usepackage{mathptmx}
\usepackage{graphicx}
\usepackage{color}
\makeindex

\begin{document}

\title{Application of superconducting-superfluid magnetohydrodynamics to nuclear ``pasta" in neutron stars}
\author{D. N. Kobyakov}
\email{dmitry.kobyakov@appl.sci-nnov.ru}
\affiliation{Institute of Applied Physics of the Russian Academy of Sciences, 603950 Nizhny Novgorod, Russia}
\date{\today}

\begin{abstract}
A mixture of superconducting and superfluid nuclear liquids of protons coupled to the ultrarelativistic electron gas, and neutrons is considered.
In the magnetohydrodynamic (MHD) approximation, the energy-momentum (stress) tensor is derived, and {the entrainment contribution is found in the explicit form.
It is shown that this contribution generates a force density, when a superfluid velocity lag and the magnetic field are simultaneously present.
This force may be important in the nuclear ``pasta" phase in neutron stars, if the proton and neutron Cooper pairing in the pasta phase is taken into account.
It is found that if the liquid-crystalline matter of the pasta phase is superfluid and superconducting, then magnitude of the forces acting upon element of matter at typical magnetic field and the superfluid velocity lag, under certain conditions may become large enough to induce a critical stress in the neutron star crust.
As an application, the necessary conditions for triggering of a starquake are found in the pasta phase of neutron stars, assuming that the nuclei are flat slabs in parallel magnetic field.
The present model includes two independent local parameters: the superfluid velocity lag and the magnetic field.}
Possible links between the entrainment force and the magnetar starquake triggering mechanism, and some open problems are discussed.
\end{abstract}
\maketitle

\section{Introduction}
Neutron star, essentially a giant atom with the mass number $\sim10^{57}$, represents a unique cosmic laboratory of dense matter.
It is expected that as the baryon density increases, the dripping of neutrons from the nuclei is followed by increase of the size of the spherical nuclei, and by joining and merging of the neighbouring nuclei, due to the forces arising from the Coulomb and the nuclear surface energies, forming lower-dimensional lattices {of the liquid-crystalline nature}, usually called ``pasta phases".
The existence of the lower-dimensional lattices is expected \cite{RavenhallPethickWilson1983,HashimotoSekiYamada1984} in the density range around one half of the symmetric nuclear saturation density {(two- and one-dimensional lattices may be possible around densities $\sim0.04$ fm$^{-3}$, and $\sim0.08$ fm$^{-3}$, correspondingly)} and has been actively explored within various models, see \cite{OyamatsuYamada1994,DouchinHaensel2000,WatanabeIidaSato2000,MagierskiHeenen2002,Horowitz2005,Maruyama2005,OyamatsuIida2007,DucoinEtal2008,GogeleinEtal2008,AvanciniEtal2010,NakazatoEtal2011,PaisStone2012,Okamoto2013,Schuetrumpf2015,SharmaEtal2015,VinasEtal2017,FattoyevHorowitzSchuetrumpf2017,LimHolt2017,NandiSchramm2018} and references therein.
One of the crucial problems is the structure and properties of the crust of neutron stars, because many observable phenomena originate at, or tightly linked to it.
Observations of the magnetic field, spin evolution, bursts, flares and some rotational properties in soft gamma repeaters and anomalous X-ray pulsars provide a base for modelling of effects of the stellar magnetic field \cite{LanderEtal2015,PassamontiPons2016}, and may be used to constrain the internal nuclear structure of neutron star crusts \cite{RavenhallPethick1994,Gearheart2011,Sotani2012,Netwon2013,PonsViganoRea2013,HorowitzEtal2015}.
Non-relativistic treatment of the elastic forces in the pasta phases was developed in \cite{PethickPotekhin1998,KobyakovPethick2018,DurelUrban2018}, and the relativistic treatment can be found in \cite{CarterSamuelsson2006}; here, the focus is on a non-relativistic model in the spirit of \cite{Mendell1991,Sedrakian1995,KobyakovPethick2017,KobyakovPethickReddySchwenk2017}.

Usually, transport of the electric charge in the pasta phases is assumed to be due to normal quasiparticles \cite{PastoreEtal2013,Yakovlev2015,SchneiderEtal2016}.
In the present paper I shall consider a physical situation when the nucleons are paired, and as a result, the nuclear matter in the pasta phase is superconducting and superfluid.
Superconductivity and superfluidity, if present, would have a strong impact for observable properties of neutron stars.
For example, evolution of the magnetic field would be influenced by a multiply-connected superconducting structure, that naturally occurs in superconducting pasta phases, because the stellar magnetic field would be trapped by persistent currents in the pasta phase; however, a resistivity below the critical superconducting temperature may be provided by the phase fluctuations leading to spontaneous appearance of {quantized vortices and magnetization} in lower-dimensional superconducting structures \cite{BlatterEtal1994}.
Another problem regards the electromagnetic stresses in the crust and generation of starquakes.
Earlier works have considered changes of the magnetic field leading to crustal stresses \cite{Lander2013,LanderEtal2015,PalapanidisEtal2015}.
Here, I shall investigate a different scenario: the nucleon velocity lag is perturbed in a constant magnetic field.
This situation is natural due to braking of rotation of the neutron star crust, that produces a superfluid velocity lag $\mathbf{w}$.

As a result of simultaneous presence of $\mathbf{w}$ and the magnetic field, the superfluid entrainment contribution to the energy-momentum tensor generates an "entrainment force".
Since this force is effective in the regions inside the mixture, where the magnetic field has penetrated, one may ask whether there may be regions in neutron stars, where the magnetic field penetrates deep enough into the bulk superconducting stellar matter, so that the entrainment force magnitude is significantly large?
In case the force is large enough, the resulting crustal stress reaches the critical value, triggering a starquake.
From the one hand, the London field in a rotating superconducting-superfluid mixture extends over the entire mixture, however, its' magnitude is small;
from the other hand, the magnetic field penetrates into a mixture with the London penetration depth, which is microscopically small for a uniform mixture \cite{AlparLangerSauls1984}.
However, in \emph{nonuniform} superconductors the effective penetration depth of the magnetic field can be quite different from the uniform case, as for instance occurs in layered superconducting materials well-known in condensed matter \cite{Tinkham1996}.

A layered superconductor is formed by an array of parallel superconducting slabs separated by non-superconducting planar regions, forming extended Josephson junctions and supporting tunneling of the Cooper pairs between the superconducting slabs.
In an extended Josephson junction there is a new characteristic length -- the Josephson penetration depth, which characterises penetration of a parallel magnetic field into the junction and depends on the strength of the Josephson tunneling.
The critical Josephson current depends on the material and may be sufficiently small, so that the Josephson penetration depth may become larger than the sample size and acquire a macroscopic value, making the magnetic screening by the Josephson currents negligible.
If, moreover, width of a single superconducting slab is smaller than the London penetration depth, then a parallel magnetic field may penetrate the entire sample of a layered superconductor.
Although the superconductivity in layered superconductors is due to the electrons, while in nuclear matter it is due to protons and the electrons are normal, structure of layered superconductors resembles that expected in the lasagna phase of nuclear matter, under assumption that the nuclear slabs are locally planar.
For simplicity, I shall focus on the lasagna phase, however other lower-dimensional structures are expected in the pasta phase, and those should be included in more realistic models of the pasta structure in the crust of neutron stars.

Using this analogy, I shall consider a toy model of lasagna phase in parallel uniform magnetic field and evaluate possible values of the entrainment force in presence of $\mathbf{w}$.
For the purpose of this paper it is sufficient to focus on the magnetic forces and to forget about the pasta elasticity.
In the simplest case, the Josephson tunneling between the slabs may be neglected, and matter inside each of the slabs would be described by the three-dimensional superconducting hydrodynamics with one of the coordinates fixed.
Therefore, as a first step, I shall develop magnetohydrodynamic formalism for uniform mixtures with the superfluid entrainment in explicit form, and then apply it to the lasagna phase in order to evaluate possible magnitude of the entrainment force.
Although the explicit form of the energy-momentum tensor has been already mentioned in the literature (see \cite{Chamel2017}), here further steps are made towards application of the MHD to neutron stars.

The MHD in uniform superfluid mixtures is crucial for interpretation of many phenomena related to neutron stars \cite{EassonPethick1979}.
Equations of motion for the mixtures with entrainment composed of two kinds of particles of almost equal masses $m_p$ and $m_n$, with $m\equiv m_p\approx m_n$, have been formulated by Mendell \cite{Mendell1991}.
As a result of neglect of the nucleon mass difference, there is a systematic error in solutions of the equations of motion.
The error is of the order of $\mu_e/m_\alpha c^2\sim10$\%, where $\mu_e$ is the electron chemical potential, $c$ is the speed of light and $m_\alpha$ is the nucleon rest mass ($\alpha=p$ for protons and $n$ for neutrons).
This approximation may be removed for various electron many-body regimes using the linear response theory \cite{KobyakovPethickReddySchwenk2017}.
While the relativistic contributions to the total mass density provide rather small quantitative corrections \cite{footnote1}, the nucleon-nucleon interactions provide qualitatively new effects such as magnetization of neutron vortices \cite{AlparLangerSauls1984}, the mode mixing and avoided crossing \cite{KobyakovPethickReddySchwenk2017}
(still, the relativistic contributions are essential from the perspective of the gravitational interaction).
Here, the focus is on physics due to the nuclear interactions, thus the relativistic corrections are neglected to a good first approximation.

\section{Momentum equations in a mixture of ideal fluids}
For an ideal fluid with the momentum per particle $\mathbf{P}$, the Euler equation has the form
$(\partial_t+\mathbf{v}\cdot\nabla)\mathbf{P}={\mathbf{F}}$.
Here, $\mathbf{v}$ is the corresponding fluid velocity, $\mathbf{F}$ is the force per particle of the fluid, which corresponds to the standard interpretation of the Euler equation given, for instance in \cite{LL_Hydro}, where a single fluid case was considered, and the momentum (per particle mass) $\mathbf{P}/m$ and velocity $\mathbf{v}$ were not distinguished.
In a mixture, each fluid is labeled by index $\alpha$ and characterised by velocity $\mathbf{v_\alpha}$.
The momentum per particle is
\begin{equation}\label{sfmomentum}
\mathbf{p}_\alpha=\hbar\nabla\phi_\alpha,
\end{equation}
where $2\phi_\alpha$ is the superfluid phase of the order parameter of the paired nucleons.
In dynamics of a particular fluid, the other fluids are self-consistent external fields.
The influence of the fields is defined by the interaction energy of the particular fluid with the other fluids, which results in nonlinear coupling of the hydrodynamic equations describing the fluids in a mixture.
The Euler equation is still valid for any fluid in the mixture (the fluid components are labeled by the index $\alpha$),
\begin{equation}\label{Euler_alpha}
\left(\frac{\partial}{\partial t}+\mathbf{v}_\alpha\cdot\nabla\right)\mathbf{P}_\alpha={\mathbf{F}_\alpha},
\end{equation}
where $\mathbf{P}_\alpha=\{\mathbf{p}_p,\mathbf{p}_n\}$ in a neutral mixture, or $\mathbf{P}_\alpha=\{\pi_p,\mathbf{p}_n\}$ in a superconducting-superfluid uniform nuclear matter, with
\begin{equation}\label{pi_p}
{{\pi}_{p}}={\mathbf{p}_p}-e\mathbf{A}/c
\end{equation}
being the proton gauge-invariant momentum per particle, $e$ the proton charge, and $\mathbf{A}$ the vector potential.
The force per particle $\mathbf{F}_\alpha$ includes all relevant interactions of the fluid $\alpha$ with other fields and fluids.
The total fluid momentum $\mathbf{P}^{\rm tot}$ is
\begin{equation}\label{Ptot}
\mathbf{P}^{\rm tot}=\sum_\alpha n_\alpha\mathbf{P}_\alpha=\sum_\alpha m n_\alpha\mathbf{v}_\alpha\equiv\sum_\alpha m\mathbf{J}_\alpha,
\end{equation}
where $n_\alpha$ is the number density of the fluid $\alpha$.
The total force
density
acting on a fluid element of a mixture is
\begin{equation}\label{Ftot}
\mathbf{F}^{\rm tot}=\sum_\alpha n_\alpha\mathbf{F}_\alpha+n_e\mathbf{F}_e,
\end{equation}
where $\mathbf{F}_e$ is the electron force per particle.

\section{The total energy and action}
In an ideal fluid, the low-energy excitations can be described by the canonically conjugated variables
-- the phase $\phi_\alpha$ and the number density $n_\alpha$ \cite{LL1980}.
The number current ${\mathbf{J}_\alpha}=n_\alpha{\mathbf{v}}_\alpha$ is defined as
\begin{equation}\label{Jnum}
\mathbf{J}_\alpha={{\partial H_{\mathrm{tot}}^{\mathrm{matt}}}}/{\partial \mathbf{P}_\alpha}.
\end{equation}
Here, $H_{\mathrm{tot}}^{\mathrm{matt}}=H_{\mathrm{tot}}^{\mathrm{matt}}(n_\alpha,\mathbf{P}_\alpha,A_0,\mathbf{A})$ is the total energy density of matter,
\begin{eqnarray}
\nonumber
  &&H_{\rm tot}^{\rm matt}=E^\mathrm{{nuc}}_{\rm st}(n_p,n_n)+E_{\rm kin}^\mathrm{{nuc}}(n_p, n_n, {{\pi}_{p}}, \mathbf{p}_n)+E^{\rm p}_{\rm Coul}(n_p, A_0)\\
  \label{Etot}
  &&+E^{\rm e}(n_e,\mathbf{J}_e,A_0)+E^{\rm e}_{\rm Coul}(n_e, \mathbf{J}_e, A_0, \mathbf{A}).
\end{eqnarray}
The static energy density $E^\mathrm{{nuc}}_{\rm st}$ is calculated from the equation of state of nuclear matter.
Macroscopic velocities of the nuclear fluids in realistic conditions are nonrelativistic, and therefore the kinetic energy density (defined as the contribution to the total energy that depends on the momentum variables) is a quadratic form of the momenta:
\begin{equation}\label{Ekn}
E_{\rm kin}^\mathrm{{nuc}}=\frac{{{{n}_{p}}}{{{\pi}}_{p}^2}}{2{m}}+\frac{{{{{n}_{n}}}}{\mathbf{p}_{n}^{2}}}{2{m}}-\frac{{{{{n }_{np}}}}}{2m}{{\left( {{{{\pi}_{p}}}}-{{{{\mathbf{p}}_{n}}}} \right)}^{2}}.
\end{equation}
Using Eqs. (\ref{Jnum}) and (\ref{Ekn}) one finds:
\begin{eqnarray}
\label{Jp}
  \mathbf{J}_p=n_p\mathbf{v}_p=\frac{n_{pp}}{m}{{\pi}_{p}}+\frac{n_{np}}{m}{\mathbf{p}_n}, \\
  \mathbf{J}_n=n_n\mathbf{v}_n=\frac{n_{nn}}{m}{\mathbf{p}_n}+\frac{n_{np}}{m}{{\pi}_{p}},
\end{eqnarray}
with $n_{\alpha\alpha}=n_{\alpha}-n_{np}$.
The entrainment number density is
\begin{equation}\label{rhonp}
\frac{{n}_{np}}{m}=\frac{{1}}{9\pi^4}k_{\mathrm{F}n}^2k_{\mathrm{F}p}^2f_1^{np},
\end{equation}
where $f_1^{np}$ is the Landau parameter and $k_{\mathrm{F}\alpha}$ is the nucleon Fermi wavenumber \cite{BorumandJoyntKluzniak1996}.

The energy density of the ultrarelativistic degenerate electrons is $E^{\rm e}={\mu_e^4}/{4\pi^2(\hbar c)^3}$, and $\mu_e=\hbar c(3\pi^2n_e)^{1/3}$ is the ultrarelativistic electron chemical potential with $n_e$ being the electron number density.
Contribution of the electrons in electromagnetic field to the total energy is given by $E^{\rm e}_{\rm Coul}=-en_e\Phi+\frac{e}{c}\mathbf{J}_e\cdot\mathbf{A}$, where $-e\mathbf{J}_e$ is the electronic current.
The proton contribution to electromagnetic interactions comes from $E^{\rm p}_{\rm Coul}=en_p\Phi$, and from the kinetic energy density.
Here, $(A_0,\mathbf{A})$ is the electromagnetic four-potential with $A_0=c\Phi$.
The total action $S=S[n_\alpha,\phi_\alpha,\Phi,\mathbf{A}]$ is
\begin{equation}\label{S}
S=\int dt\int d^3\mathbf{r}({\mathcal{T}}-H_{\mathrm{tot}}^{\mathrm{matt}})+S^{\mathrm{em}},
\end{equation}
where $\mathcal{T}=-\hbar n_{p}{\partial_t{\phi}}_{p}-\hbar n_{n}\partial_t{\phi}_{n}$ and $S^{\mathrm{em}}=({1}/{{16\pi}})\int dt\int d^3\mathbf{r}\sum_{a,b=0,1,2,3}(\partial_aA_b-\partial_bA_a)^2$.
It is interesting to notice, and straightforward to check by calculations, that the dynamic equations based on the action in Eq. (\ref{S}) can be precisely mapped into the Schr$\mathrm{\ddot{o}}$dinger model with a modified quantum pressure term.
Therefore, the non-trivial forces acting upon the vortices, such as the Magnus force, are implicitly and automatically included in the equations of motion.
The forces can be found by integrating out the irrelevant degrees of freedom in the superfluid action (for example, see \cite{KobyakovEtal2012}, Sec. IV.B).

\section{The superfluid equations of motion}
The equations of motion are obtained from Eq. (\ref{S}) by the variational method, see \emph{e. g.} \cite{DubrovinNovikovFomenko1984}.
The Euler-Lagrange equations for the fields $n_\alpha$ and $\phi_\alpha$ are
\begin{eqnarray}\label{contalpha}
\partial_t n_\alpha + \nabla\cdot\mathbf{J}_\alpha=0,\\
\label{Josalpha}
\partial_t \phi_\alpha + {\mu_{\alpha}^{\rm tot}}/{\hbar}=0.
\end{eqnarray}
Equations (\ref{Josalpha}) are well-known in the context of superconducting mixtures with entrainment \cite{Mendell1991,Sedrakian1995,Mendell1998,KobyakovPethick2017}.
The total chemical potentials defined as
\begin{equation}\label{mualphatot}
\mu_{\alpha}^{\rm tot}=\partial H_{\rm tot}^{\rm matt}/\partial n_\alpha,
\end{equation}
can be easily calculated using Eqs. (\ref{Etot}) and (\ref{Ekn}):
\begin{eqnarray}\label{muptot}
  \mu_{p}^{\rm tot}=\frac{\pi_p^2}{2m}-\frac{m}{2}\frac{\partial n_{np}}{\partial n_{p}}\mathbf{w}^2+\mu_p^{\rm nuc}+e\Phi,\\
  \label{muntot}
  \mu_{n}^{\rm tot}=\frac{\mathbf{p}_n^2}{2m}-\frac{m}{2}\frac{\partial n_{np}}{\partial n_{n}}\mathbf{w}^2+\mu_n^{\rm nuc},
\end{eqnarray}
where
\begin{equation}\label{mualphanuc}
\mu_\alpha^\mathrm{{nuc}}={\partial E^\mathrm{{nuc}}_{\rm st}}/{\partial n_\alpha}
\end{equation}
are \emph{the chemical potentials}.
I should like to emphasize that in Eq. (\ref{mualphanuc}) the chemical potentials are defined in the absence of macroscopic nucleon flows ($\mathbf{v}_\alpha=0$).
The momentum lag is
\begin{equation}\label{w}
\mathbf{w}=\pi_p/m-\mathbf{p}_n/m.
\end{equation}
Here, the gravitational potential is not explicitly included in the equations of motion, however, a generalization is straightforward.
The explicit form of chemical potentials in Eqs. (\ref{muptot}) and (\ref{muntot}) is well-known
in the context of superfluid mixtures with entrainment \cite{Mendell1991,KobyakovPethick2017,Recati2017}.
The equations for the nucleon momenta are obtained by application of $\nabla$ to both sides of Eq. (\ref{Josalpha}),
\begin{equation}\label{MomentaEquations}
\partial_t\mathbf{p}_\alpha+\nabla\mu^{\rm tot}_\alpha=0,
\end{equation}
or equivalently, using $(1/2)\nabla\mathrm{\mathbf{w}}^2\equiv(\mathrm{\mathbf{w}}\cdot\nabla)\mathrm{\mathbf{w}}+\mathrm{\mathbf{w}}\times\nabla\times\mathrm{\mathbf{w}}$:
\begin{eqnarray}
\label{momentumEq}
&&\left(\frac{\partial}{\partial t}+\mathbf{v}_p\cdot\nabla\right)\pi_p={\mathbf{F}_p},\quad
\left(\frac{\partial}{\partial t}+\mathbf{v}_n\cdot\nabla\right)\mathbf{p}_n={\mathbf{F}_n},\\
\nonumber
&&\mathbf{F}_p=-\nabla\mu_p^{\rm nuc}+e\mathbf{E}+\frac{e}{c}\mathbf{v}_p\times\mathbf{B} -\frac{\partial n_{np}}{\partial n_p}(\mathbf{w}\cdot\nabla)\mathbf{p}_n\\
  \label{ForceP}
&&+\frac{m\mathbf{w}^2}{2}\nabla\frac{\partial n_{np}}{\partial n_p}-\theta_p\left[(\mathbf{w}\cdot\nabla)\pi_p-\frac{e}{c}\mathbf{w}\times\mathbf{B}\right]\\
\nonumber
&&\mathbf{F}_n=-\nabla\mu_n^{\rm nuc}+\frac{\partial n_{np}}{\partial n_n}\left[(\mathbf{w}\cdot\nabla)\pi_p-\frac{e}{c}\mathbf{w}\times\mathbf{B}\right]\\
  \label{ForceN}
&&+\frac{m\mathbf{w}^2}{2}\nabla\frac{\partial n_{np}}{\partial n_n}+\theta_n(\mathbf{w}\cdot\nabla)\mathbf{p}_n,
\end{eqnarray}
where $\mathbf{E}=c^{-1}(-{\nabla} A_0-\partial_t\mathbf{A})$ is the electric field and $\mathbf{B}=\nabla\times\mathbf{A}$ is the magnetic induction.
The forces defined in Eqs. (\ref{ForceP}) and (\ref{ForceN}) can be interpreted as following.
The first terms are the usual thermodynamic contributions; the second and third terms in $\mathbf{F}_p$ are the Lorentz force; the fifth term in $\mathbf{F}_p$ and the third term of the sum in $\mathbf{F}_n$ are the surface forces due to the entrainment; the second term in $\mathbf{F}_n$ contains the neutron magnetic force density $-(e/c)({\partial n_{np}}/{\partial n_n})\mathbf{w}\times\mathbf{B}$.
The nuclear interaction corrections are given by the functions
\begin{equation}\label{theta}
\theta_\alpha=\frac{n_{np}}{n_\alpha}-\frac{\partial n_{np}}{\partial n_\alpha}.
\end{equation}
In the mean field models of nuclear matter one usually assumes $\theta_\alpha=0$ \cite{KobyakovPethickReddySchwenk2017}.

Variation of $S$ with respect to the four-vector potential $(A_0,\mathbf{A})$ leads to Maxwell equations
\begin{eqnarray}
  \label{Maxwell1}
  {\nabla}\times({\nabla}\times \mathbf{A})-\frac{1}{c}\partial_t\mathbf{E}=\frac{4\pi}{c}e\mathbf{J}_{\rm tot},\\
  \label{Maxwell2}
  {\nabla}\cdot\mathbf{E}=4\pi e\left(n_p-n_e\right),
\end{eqnarray}
where $e\mathbf{J}_{\rm tot}=e\mathbf{J}_p-e\mathbf{J}_e$ is the total electric current.
The superconducting contribution is given by $e\mathbf{J}_p$.
The error due to the Newtonian limit of the Maxwell equations is expected to be negligible for conditions relevant to neutron star modelling \cite{Sedrakian1995,CarterLanglois1998,Prix2005}.
Using the Coulomb gauge for the vector potential, ${\nabla}\cdot\mathbf{A}=0$, from Eq. (\ref{Maxwell2}) one obtains the Poisson equation
 \begin{equation}\label{PoissonEqn}
   {\nabla}^2\Phi=-4\pi e\left(n_p-n_e\right).
 \end{equation}
The electron density in the Poisson equation can be excluded with the help of the linear response theory \cite{KobyakovPethickReddySchwenk2017}.
In the hydrodynamic regime, the frequency-dependent response function provides the electron inertia corrections in the proton momentum equation, while its' static limit leads to the non-relativistic momentum equations \cite{Kobyakov2017}.
Furthermore, neglecting $\partial_t\mathbf{A}$ in the definition of $\mathbf{E}$ and using the expression for $\Phi$ obtained from the Poisson equation, allows to exclude the electric field from Eq. (\ref{ForceP}) (the electric neutrality is maintained at long wavelengths).
Finally, neglecting the displacement current in Eq. (\ref{Maxwell1}) leads to the equation for $\mathbf{B}$ in terms of the fluid variables, and thus the set of Eqs. (\ref{contalpha}), (\ref{momentumEq}) and (\ref{Maxwell1}) is closed.

\section{Energy-momentum tensor}
The total momentum conservation reads
\begin{equation}\label{Pik_def}
  \frac{\partial}{\partial t}P_i^{\rm tot}+\sum_k\frac{\partial}{\partial x_k}\Pi_{ik}=0,
\end{equation}
with $(i,k=1,2,3)$.
From the Euler equations, Eq. (\ref{momentumEq}), and the definition Eq. (\ref{Pik_def}), one finds
\begin{equation}\label{gradPI}
\sum_k\frac{\partial}{\partial x_k}\Pi_{ik}=(\mathbf{J}_{p}\cdot\nabla)\pi_{p\,i}+(\mathbf{J}_{n}\cdot\nabla)p_{n\,i}-F_i^{\rm tot}.
\end{equation}
The electron force per particle is
\begin{equation}\label{FLe}
\mathbf{F}_{e}=-\nabla\mu_e-e\mathbf{E}-\frac{e}{c}n_e^{-1}\mathbf{J}_e\times\mathbf{B}.
\end{equation}
Using Eqs. (\ref{Ftot}), (\ref{ForceP}), (\ref{ForceN}), (\ref{Maxwell1}), (\ref{FLe}) and neglecting the displacement current, one finds the  momentum flux tensor in the superconducting superfluid mixture
\begin{eqnarray}
\nonumber
&&\Pi_{ik}=J_{p\,k}\pi_{p\,i}+J_{n\,k}p_{n\,i}+\delta_{ik}p-{\rho_{np}^*}\mathrm{w}_k\mathrm{w}_i\\
\label{PIsfin}
&&-{\rho_{np}^*}(\delta_{ik}\frac{\mathrm{\mathbf{w}}^2}{2}-\mathrm{w}_k\mathrm{w}_i)+\frac{1}{4\pi}(\delta_{ik}\frac{B^2}{2}-B_kB_i),
\end{eqnarray}
where $\rho_{np}^*=mn_{np}-mn_p\theta_p-mn_n\theta_n$.
The quantity $p=p(n_p,n_n)$ is the pressure defined in the absence of flows and the magnetic field:
\begin{equation}\label{pressure}
  p=n_e\mu_e+n_p\mu_p^{\rm nuc}+n_n\mu_n^{\rm nuc}-E^{\rm nuc}_{\rm st}-E^{\rm e}.
\end{equation}
The tensor in Eq. (\ref{PIsfin}) is implicitly equivalent to the stress tensor found for the first time in \cite{Sedrakian1995}.
Equation (\ref{PIsfin}) is explicitly equivalent to the result obtained independently in \cite{Chamel2017}, that came up shortly before appearance of the present paper \cite{Kobyakov2017}.
For the practical purposes it is more convenient to write the stress tensor working with the Helmholtz free energy \cite{EassonPethick1977}.

The results obtained above agree with the results found in the literature.
The pressure defined in the earlier works as $P=P(n_\alpha,\mathbf{w}^2)=-\mathcal{E}_{in}+n_p\mu_p^{\rm tot}+n_n\mu_n^{\rm tot}+n_e\mu_e$ \cite{Sedrakian1995}, is related to the pressure $p$ defined in the absence of flows,
\begin{equation}\label{Pp}
  P=p-\rho_{np}^{*}\frac{\mathrm{\mathbf{w}}^2}{2},
\end{equation}
as follows from Eq. (\ref{PIsfin}).
Equation (\ref{Pp}) shows that $P$ includes contribution from the velocity lag, therefore $P$ is a function of the vector potential $\mathbf{A}$, and the pressure contribution to the force $\nabla P$ may depend on the magnetic field.
Moreover, if the relativistic contributions to the total mass density are retained \cite{Kobyakov2017}, it is straightforward to find that, in fact, $P=P(n_\alpha,\pi_p^2,\mathbf{p}_n^2,\mathbf{w}^2)$.
The quantities associated with the chemical potentials in \cite{Sedrakian1995} are denoted here as the total chemical potentials, Eqs. (\ref{muptot}) and (\ref{muntot}).
One of advantages of the present formulation is that the definitions of the chemical potentials and the pressure, Eqs. (\ref{mualphanuc}) and (\ref{pressure}), do not include the superfluid momenta, and thus the thermodynamic variables are defined in the absence of matter flows.
In the present approach, the momentum dependence of the equations of motion, Eq. (\ref{momentumEq}), is explicit.

\section{A single slab in uniform magnetic field}
As a first step, a single flat slab at zero temperature is considered, parallel to $x-y$ plane and located (rigidly fixed) between the planes $z=\pm r_N$.
This resembles a classical problem of a flat slab in parallel magnetic field \cite{Tinkham1996}, but the present consideration includes also a possibility of the superfluid entrainment.
The slab is then immersed into a uniform magnetic field $\mathbf{H}$, directed along $x$ axis.
Typical slab width is $2r_N\sim10$ fm \cite{WatanabeIidaSato2000} and is much smaller than the coherence length and the London penetration depth in the uniform matter ($\sim30$ fm and $\lambda\sim80$ fm correspondingly \cite{AlparLangerSauls1984}), thus the magnetic field is approximately uniform inside a single slab.

The latter statement is proved within a classic problem of a flat slab in parallel magnetic field \cite{Tinkham1996}.
Using the boundary conditions at the surfaces of the slab at $z=\pm r_N$ that the microscopic magnetic field $\mathbf{h}$ is equal to the ambient magnetic field $\mathbf{H}$,
\begin{equation}\label{hH}
  \mathbf{h}(z=\pm r_N)=\mathbf{H},
\end{equation}
one obtains neglecting terms of the order of {$(r_N/2\lambda)^2$ and the higher-order terms}, that the screening by the Meissner currents around the slab is negligible, that is, the magnetization of the slab is zero, and the magnetic induction - the average of the microscopic field $\mathbf{h}$ over the slab width $\mathbf{B}=\mathbf{H}$ [see \cite{Tinkham1996}, equation (2.5)].
The ambient magnetic field is denoted as $\mathbf{H}=\mathbf{B}_0$, where $\mathbf{B}_0=B_0\mathbf{\hat{x}}$.

While the superfluid phase is strictly a two-dimensional function inside the superconducting domain, the vector potential conserves the microscopic character across the slab, and has to be distinguished at the boundaries of the slab.
In equilibrium, the vector potential inside the slab is $\mathbf{A}_0\approx-\mathbf{\hat{y}}B_0z$.
The unperturbed superfluid gauge-invariant momentum lag reads
\begin{equation}\label{w0z}
\mathbf{w}_0(z)=-(e/mc)\mathbf{A}_0,
\end{equation}
or $\mathbf{w}_0(z)=(e/mc)\mathbf{\hat{y}}B_0z$.
As usually, the normal component of the superconductor is at rest in the laboratory frame of reference (defined as the rest frame for the normal charged component of the superconductor), therefore the proton phase is spatially constant inside the slab in equilibrium, and the equilibrium value of the superfluid proton momentum inside the slab centered at $z=0$ is zero, $p_{p0}=0$, and is dropped in Eq. (\ref{w0z}).
The nucleon densities are approximated as following:
\begin{eqnarray}
  \label{n_p_z}
  &&n_p(z)=n_{p0}\left[\theta(z+r_N)-\theta(z-r_N)\right], \\
  \label{n_n_z}
  &&n_n(z)=n_{n0}^{o}+\left(n_{n0}-n_{n0}^{o}\right)\left[\theta(z+r_N)-\theta(z-r_N)\right],
\end{eqnarray}
where $\theta(z)$ is the Heaviside function, $n_{\alpha0}$ are nucleon densities inside the slab, and $n_{n0}^{o}$ is the density of neutron matter outside the slab.

A perturbation of the superfluid momentum lag with a uniform spatial distribution
\begin{equation}\label{dw}
\delta\mathbf{w}=\mathbf{\hat{y}}\delta\mathrm{w},
\end{equation}
is imposed on top of the background nucleon densities, phase gradients $\mathbf{p}_{\alpha0}$, and the vector potential $\mathbf{A}_0$.
{The entrainment contribution is a function of an independent parameter - the nucleon velocity lag, therefore in a linear analysis, one may assume that the} superconducting current is unperturbed,
\begin{equation}\label{deltaJp}
\delta\mathbf{J}_p=0.
\end{equation}
The total force acting on 1 cm$^2$ of a single slab is found with the help of linearization of the total force density to the first order in $\delta\rm{w}$, assumption $\theta_\alpha=0$, the use of Eqs. (\ref{Ftot}), (\ref{ForceP}), (\ref{ForceN}), and (\ref{n_p_z})-(\ref{deltaJp}), and integration over space:
\begin{eqnarray}
\nonumber
  &&\delta^{(1)}\mathbf{F}_{\rm tot}^{1{\,\rm cm}^2\times 2r_c}=1{\,\rm cm}^2\times\int_{-r_c}^{r_c}dz \:\left[\right. m\delta\mathbf{w}\cdot\mathbf{w}_0(z)\\
  \nonumber
  &&\times\left(n_{np0}\frac{n_{n0}-n_{n0}^o}{n_{n0}}+n_{np0}\right)\mathbf{e_z}\left(\delta\left(z+r_N\right)-\delta\left(z-r_N\right)\right)\\
  \nonumber
  &&-\frac{e}{c}n_{np}(z)\delta\mathbf{w}\times\mathbf{B}_0\left.\right]\\
  \label{delta1Ftot}
  &&=-1{\,\rm cm}^2\times\mathbf{\hat{z}}\frac{e}{c}B_0\,2r_N\,n_{np0}\frac{n_{n0}-n_{n0}^o}{n_{n0}}\delta{\rm w},
\end{eqnarray}
where $r_c$ is the radius of the unit cell in the lattice, and $n_{np0}$ corresponds to the nuclear density inside the slab.
It is interesting to note that the three terms integrated in Eq. (\ref{delta1Ftot}) can be viewed as the nucleon surface terms $\propto\delta(z\pm r_N)$, and the volume term solely due to neutrons (the last term).

\section{Towards astrophysical applications}
The consideration of a single slab can be generalized to the case of a single-dimensional array of slabs immersed in a parallel magnetic field.
The Josephson critical current is a phenomenological parameter, and, unlike in the condensed matter applications, in nuclear matter it cannot be found straightforwardly.
In fact, the ground state of the pasta phase is an open problem.
Here, for simplicity, it is assumed that the slabs are planar, and that the Josephson penetration depth is macroscopically large, so tunneling between the slabs is neglected.
In this toy model, a uniform $\mathbf{B}=\mathbf{B}_0$ is along $x$ axis and is between $z=-d/2$ and $z=d/2$ planes, and penetrates the entire sample of the lasagna matter.
This is in accordance with the fact that the width of a single slab is much smaller than the London penetration depth:
using the boundary conditions given in Eq. (\ref{hH}) for each of the slabs, one obtains solutions that represent uniformly penetrating parallel magnetic field with the negligible screening by the Josephson currents, and with a macroscopically large Josephson penetration depth.
Another important assumption is that the superfluid velocity lag is perpendicular to the magnetic field and parallel to the slab surface.
These conditions are rather specific, however it cannot be excluded that the conditions hold in some regions in the crusts of neutron stars due to presence of a toroidal component of the magnetic field {(for a recent discussion of the toroidal magnetic field in neutron stars, see \cite{LanderJones2018})}, and also because spatial orientation of the anisotropic matter in the nuclear pasta phase is uncertain.

Since the focus is on evaluation of possible magnitude of the entrainment force, the lattice is assumed to be rigidly fixed and at rest.
In this case, the number current densities can be written as $\mathbf{J}_p=\hat{\hat{n}}_{pp}\pi_{p}+\hat{\hat{n}}_{np}\mathbf{p}_{n}$ and $\mathbf{J}_n=\hat{\hat{n}}_{nn}\mathbf{p}_{n}+\hat{\hat{n}}_{np}\pi_{p}$, where the tensors $\hat{\hat{n}}_{\alpha\beta}$ represent the anisotropic superconducting density, and the structure is fixed, $\dot{\mathbf{\rm u}}=0$, where ${\mathbf{\rm u}}$ is the displacement vector of the lattice (see \cite{KobyakovPethick2018}).
In one-dimensional lattice of nuclei, with the slabs parallel to $x-y$ plane, the proton current is $\mathbf{J}_p={n}_{pp}^{\perp}{\pi_{p}}_z\mathbf{\hat{z}}+{n}_{pp}^{\parallel}({\pi_{p}}_x\mathbf{\hat{x}}+{\pi_{p}}_y\mathbf{\hat{y}})+{n}_{np}^{\perp}{{p}_{n}}_z\mathbf{\hat{z}}+{n}_{np}^{\parallel}({{p}_{n}}_x\mathbf{\hat{x}}+{{p}_{n}}_y\mathbf{\hat{y}})$.
Since tunneling between the layers is assumed to be negligible, then ${n}_{pp}^{\perp}={n}_{np}^{\perp}=0$.

The equilibrium value of the proton superfluid phase $\mathbf{p}_{p0}$ is by initial conditions chosen at each sheet in order to cancel the contribution from $\mathbf{A}_0$: $\mathbf{w}_{0}(x,y,z_{s})=\mathbf{p}_{p0}(z_{s})/m-(e/mc)\mathbf{A}_0(z_{s})-\mathbf{p}_{n0}/m$, where $z_s$ is the position of the middle of a slab.
The equilibrium momentum lag is zero in the middle of each slab,
\begin{equation}\label{w0}
\mathbf{w}_0(x,y,z)|_{z=z_s}=0.
\end{equation}
Typical separation between the slabs is $2r_c\sim20$ fm \cite{WatanabeIidaSato2000} (see \cite{CaplanHorowitz2017} for a recent review).
If the coherence length were much smaller than separation between the slabs, one would simply choose the phase at each slab so that the condition in Eq. (\ref{w0}) is satisfied, but here, the two lengths are of the same order of magnitude.
Thus, in order to provide the necessary phase winding that cancels the contribution to the proton current density from $\mathbf{A}_0(z)$ in the middle of each slab in Eq. (\ref{w0}), the Josephson vortices are introduced into the voids between the slabs.
The Josephson vortices differ from the usual magnetic fluxtubes, because they exist in the region where the superconducting density is zero, while the superconductor is residing in a multiply-connected domain.

Before a starquake, the fluid element is in equilibrium and does not move along $z$, according to the initial assumption that the structure is fixed.
By virtue of continuity of stress at the crust-lasagna boundary, the entrainment force exerts an external elastic stress on the crust, $\sigma^{\mathrm{ex}}_{ij}$, which consists of a single nonzero $zz$ component,  $\sigma^{\mathrm{ex}}_{zz}=(\delta\mathbf{F}_{n}^{1{\rm cm}^2\times d})_z$.
The stress balance in the solid crust reads:
\begin{equation}\label{balance}
0=-p\delta_{ij}+\sigma^{\mathrm{ex}}_{ij}+\mu_{\mathrm{eff}} u_{ij}+M_{ij},
\end{equation}
where $p$ is the pressure, $\sigma^{\mathrm{ex}}_{ij}$ is the external stress, $\mu_{\mathrm{eff}}$ is the effective shear modulus of polycrystalline solid at the bottom of the inner crust with spherical nuclei, and $M_{ij}$ is the Maxwell stress tensor.
Thus, the induced strain $u_{ij}$ due to the perturbation $\delta\mathbf{w}$ is
\begin{equation}\label{uij}
\mu_{\mathrm{eff}} u_{ij}=-\left(\delta\mathbf{F}_{\rm tot}^{1{\rm cm}^2\times d}\right)_z.
\end{equation}
In order to estimate $(\delta\mathbf{F}_{n}^{1{\rm cm}^2\times d})_z$, it is assumed that the baryon density inside the slab is 0.16 ${\rm fm}^{-3}$, and thus, $n_{np}\sim-3.692\times10^{-3}$ fm$^{-3}$ \cite{KobyakovPethickReddySchwenk2017}.
The perturbation of the velocity lag $\mathbf{v_p}-\mathbf{v_n}$ which is assumed to be $\sim1$ cm s$^{-1}$ in typical conditions, around its' equilibrium zero value, is related to $\delta\mathbf{w}$: $\mathbf{v_p}-\mathbf{v_n}=(n_pn_n)^{-1}(\det{n_{\alpha\beta}})\delta\mathbf{w}$, and one finds $\mathbf{v_p}-\mathbf{v_n}=1.501\delta\mathbf{w}$.
For the order of magnitude estimates it is assumed that $(n_{n0}-n_{n0}^o)/n_{n0}\sim1$, which is expected to provide error of the order of 50\%.
The total force acting on a $1{\rm cm}^2\times d$ column of lasagna immersed in a uniform magnetic field is
\begin{eqnarray}
\nonumber
&&\left|\left(\delta\mathbf{F}_{\rm tot}^{1{\rm cm}^2\times d}\right)_z\right|\sim \frac{1\,\mathrm{cm}}{2r_c}\left|\delta^{(1)}\mathbf{F}_{\rm tot}^{1{\,\rm cm}^2\times 2r_c}\right|=1.971\times10^{30}\\
\label{Fln1d}
&& \times\left(\frac{d}{1\,\mathrm{cm}}\right)\left(\frac{\delta\mathrm{w}}{1\,\mathrm{cm\,s^{-1}}}\right)\left(\frac{B}{10^{14}\,\mathrm{G}}\right)\;[\mathrm{dyn}],
\end{eqnarray}
where $d$ is the penetration depth along $z$ of the magnetic field that is parallel to the slab surface.
This estimate is easy to understand because the contribution from a single slab, Eq. (\ref{delta1Ftot}), is of the order of $10^{18}$ dyn, and there are about $(d/1\,\rm cm)\times10^{12}$ slabs in a 1 cm$^2\times d$ column of lasagna matter.
The typical scale of $d$ in the engineering formula in Eq. (\ref{Fln1d}) must be corrected upon solution of the problem of penetration of the magnetic field into a realistic configuration of the pasta phase.
It is also important to investigate the effect of stratification in the pasta phase on the effective penetration depth of the magnetic field.
Notably, calculations searching for the ground state of the pasta phase require very high precision in order to distinguish between the true ground state in the global minimum of the free energy, and the local minima.
It is also necessary to take into account variations of the angle between the magnetic field and the pasta anisotropy directions, however, this goes beyond the scope of this paper.
Here, the force arising from the isotropic part of the stress tensor was considered.
This force was neglected in \cite{EassonPethick1977} as a result of neglect of the dependence of the fluid free energy on the fluid density.

The crust yielding occurs when von Mises criterion is satisfied, $\sqrt{u_{ij}u_{ij}/2}\geq u_{max}$ \cite{LanderEtal2015}.
The neutron star crusts are expected to be polycrystalline \cite{KobyakovPethick2015,CaplanEtal2018}, but further work is clearly necessary for better understanding of the structure.
Here, I assume that the effective shear modulus of the crust is described by the averaged result valid for polycrystalline solids, $\mu_{\mathrm{eff}}=0.3778\frac{n_NZ^2e^2}{2a}$ \cite{KobyakovPethick2015}.
I use the parameters at baryon density $7.943\times10^{-2}$ \cite{KobyakovPethick2016}: $n_N=1.750\times10^{-4}$ fm$^{-3}$, $Z=17.23$, $a=(3/4\pi n_N)^{1/3}=11.09$ fm, and using Eq. (\ref{uij}) {find} that the crust yields when
\begin{equation}\label{vonMises}
\frac{\left|\left(\delta\mathbf{F}_{\rm tot}^{1{\rm cm}^2\times d}\right)_z\right|}{2.040 \times10^{30}\,\mathrm{dyn}}\geq \sqrt{2}u_{max}.
\end{equation}
For $u_{\rm max}\sim0.1$ \cite{HorowitzKadau2009}, typical entrainment force estimated in Eq. (\ref{Fln1d}), is a few times larger than the critical stress that breaks the crust.
In the present model with typical magnetic field $10^{14}$ G, {the decoupling of the Lorentz force from the analysis is in agreement with the prediction of \cite{LanderEtal2015} that the crust breaking at changing magnetic field and zero superfluid momenta lag $\mathbf{w}=0$, with the critical strain of the solid crust} $u_{\rm max}\sim0.1$ \cite{HorowitzKadau2009}, occurs at $2.4\times10^{15}$ G.

The precise x-ray burst mechanism in magnetars is an open question \cite{TurollaZaneWatts2015,KaspiBeloborodov2017}.
If the bursts are generated by starquakes, it seems possible that the entrainment force is capable to explain triggering of starquakes.
An important problem for future study is characterization of structure of the nuclear matter inside neutron stars, in particular, determination of stratified superconducting  and superfluid density profiles, sizes and orientations of the crystalline domains in the magnetic field, and better understanding of the elastic, magnetic and transport properties of the pasta phases.

\section{Acknowledgments}
Part of this work has been done while I enjoyed the hospitality of the Ioffe Physical Technical Institute in Saint Petersburg.
I thank the anonymous referee for useful remarks.
This work was supported by the Russian Foundation for Basic Research, according to the research project No. 31 16-32-60023 mol$_-$a$_-$dk.

\appendix
\section{Comparison with earlier work}
Equations describing the superfluid mixture in the core of neutron stars were also studied in the framework of the convective variational approach \cite{CarterKhalatnikov1992,PrixComerAndersson2002,Prix2004,ACP2004}.
A subtle, yet crucial detail that distinguishes the convective formulation from the canonical Hamiltonian formulation is the difference between the quantity $\mu_X$ (which was denoted $\mu_X^{\mathrm{ACP}}$ in \cite{KobyakovPethick2017}) and the usual thermodynamic chemical potential $\mu_\alpha^{\rm nuc}$.
Calculation of partial derivatives according to equation (66) in \cite{Prix2004} leads to
\begin{equation}\label{muACP}
  \mu_X^{\mathrm{ACP}}=\mu_X^{\mathrm{nuc}}+\frac{\partial \alpha^{XY}}{\partial n_X}(\mathbf{v}_X-\mathbf{v}_Y)^2,
\end{equation}
where the subscripts $\{X,Y\}$ correspond to $\{\alpha,\beta\}$, and $\alpha^{XY}=-(m/2)n_{np}n_pn_n/\det{n_{\alpha\beta}}$.
Thus, the quantity $\mu_X^{\mathrm{ACP}}$ has an implicit dependence on the velocity lag, similarly to the quantity $P$ discussed above.

It is instructive to show, that with the definition Eq. (\ref{muACP}) the equations of motion of the convective approach are equivalent to the Hamilton equations derived here, Eq. (\ref{Josalpha}).
Forgetting for a moment about the temperature gradient and the mutual friction forces, one can cast equations (176) and (177) in \cite{Prix2004} to the form
\begin{eqnarray}
\nonumber
&&\frac{\partial}{\partial t}\frac{{{\mathbf p}_{X}}}{m}={{\mathbf v}_{X}}\times\nabla\times\frac{{{\mathbf p}_{X}}}{m}-\nabla \left(\frac{\mathbf{p}_{X}^{2}}{2{{m}^{2}}}-\frac{\epsilon_{X}^{2}{({{{\mathbf v}_{X}-{\mathbf v}_{Y}}})^{2}}}{2}+\frac{\mu_{X}^{\mathrm{{ACP}}}}{m}\right),\\
\label{ACPeqs}
\end{eqnarray}
where $\epsilon_{X}=2\alpha^{np}/mn_X$.
The gradient term in Eq. (\ref{ACPeqs}) is equivalent to the gradient term in Eq. (\ref{MomentaEquations}).
The first term in the right side of Eq. (\ref{ACPeqs}) disappears inside electrically neutral superfluids.

To the best of my knowledge, Eq. (\ref{muACP}) has not appeared in the earlier literature.
{As has been discussed in \cite{KobyakovPethick2017}, this fact has hindered comparison of the two approaches to the problem of the collective modes in a uniform superfluid mixture with a superfluid counterflow developed in \cite{ACP2004} and in \cite{KobyakovPethick2017}.
Here, using the definition in Eq. (\ref{muACP}), equivalence of the equations of motion, and absence of a conflict, between the two approaches has been demonstrated explicitly in Eq. (\ref{ACPeqs}).
In fact, the linear analysis in \cite{ACP2004} has been done in a model without the entrainment, however, the basic formalism is equivalent to the Hamilton equations used here.}
The dispersion relation for the coupled modes in the general case when the entrainment and the superfluid velocity lag are taken into account, has been found in \cite{KobyakovPethick2017}.

The convective approach has been generalized to the case of superconducting fluid mixture in \cite{CarterLanglois1998,Prix2005,GlampedakisAnderssonSamuelsson2011}.
When a superfluid is electrically charged, the first term in the right side of Eq. (\ref{ACPeqs}) may be non-zero inside the superconducting bulk, but the additional force exactly balances this term, and equation (30) in \cite{Prix2005} [or equation (29) in \cite{GlampedakisAnderssonSamuelsson2011}] becomes explicitly equivalent to Eq. (\ref{MomentaEquations}).

\end{document}